\documentclass{egpubl}
\usepackage{vcbm2024p}

\WsPoster          

\usepackage[T1]{fontenc}
\usepackage{dfadobe}

\usepackage{cite}  
\BibtexOrBiblatex
\electronicVersion
\PrintedOrElectronic
\ifpdf \usepackage[pdftex]{graphicx} \pdfcompresslevel=9
\else \usepackage[dvips]{graphicx} \fi

\usepackage{egweblnk}

\title[CaroTo: A Tool for Fast Comprehensive Analysis of Carotid Artery Stenosis]%
      {CaroTo: A Tool for Fast Comprehensive Analysis of Carotid Artery Stenosis in 4D PC- and 3D BB-MRI Data}

\author[H. Rahlfs et al.]
{\parbox{\textwidth}{\centering H. Rahlfs$^{1}$\orcid{0009-0005-1870-9290}, M. Hüllebrand $^{1, 2, 3}$\orcid{0000-0003-4948-0917}, S. Schmitter$^{4}$\orcid{0000-0003-4410-6790}, J. Andrae$^{5}$, C. Strecker$^{5}$, A. Harloff$^{5}$\orcid{0000-0002-3252-7910}, and A. Hennemuth$^{1,2,3}$\orcid{0000-0002-0737-7375} 
        }
        \\
{\parbox{\textwidth}{\centering $^1$ Institute of Computer-Assisted Cardiovascular Medicine, Charité - Universitätsmedizin Berlin, Germany\\
         $^2$ Fraunhofer MEVIS, Bremen, Germany\\
         $^3$ DZHK (German Centre for Cardiovascular Research), Partner Site Berlin, Germany\\
         $^4$ Physikalisch-Technische Bundesanstalt, Berlin, Germany\\
         $^5$ Department of Neurology and Neurophysiology, Faculty of Medicine, Medical Center—University of Freiburg, University of Freiburg, Germany
       }
}
}

\begin{document}


\maketitle
\begin{abstract}
   Atherosclerosis of the carotid artery increases stroke risk. Atherosclerosis assessment with MRI requires multimodal and multidimensional segmentation of the carotid artery, reproducible extraction of biomarkers, and the visualization of segmentations and biomarkers. We developed CaroTo, a tool that allows for standardized carotid atherosclerosis assessment. It combines the capabilities of MEVISFlow with  specialized tools for carotid geometry and vessel wall assessment. It supports manual and automatic segmentation for 2D, 2D+time, and 3D images, facilitating precise and consistent evaluations of carotid artery stenosis.


\end{abstract}  
\section{Introduction}

Atherosclerosis of the carotid artery is a major risk factor for stroke, a leading cause of disability and death. \cite{CDM*19} It can be assessed with magnetic resonance imaging (MRI). This includes Time-of-Flight (ToF)-MRI for assessment of the lumen geometry and stenosis degree, 3D black blood (BB)-MRI for the assessment of wall thickness and plaque properties, and 3D + time phase contrast (PC)-MRI, also called 4D-Flow MRI, for the assessment of the hemodynamics. 

The reproducible extraction of biomarkers like vessel wall thickness (VWT), wall shear stress (WSS), and pulse wave velocity (PWV) requires a segmentation of the carotid artery wall in cross-sections that are perpendicular to the centerline and at standardized positions \cite{SKK*20}. The segmentation can be done manually or by using automatic segmentation with manual correction. The development of automatic segmentation methods for 3D and 4D MRI data promises more expressive and reproducible analysis of carotid artery stenosis but requires proper visualization for annotation correction and visual assessment of the multidimensional data and extracted features.

\section{CaroTo}

We developed CaroTo, a tool for the analysis of carotid artery stenosis. It combines functionalities of MEVISFlow for the segmentation and visualization of 4D-Flow MRI and a carotid artery specific geometry analysis and vessel wall assessment on BB- and ToF-MRI. Currently, CaroTo allows for the manual segmentation of 2D, 2D + time, and 3D segmentation of the carotid artery. Additionally, manual correction of automatic 2D and 3D segmentation is possible for BB-MRI.

\subsection{Segmentation}
\begin{figure*}[tbp]
  \centering
  \includegraphics[width=1\linewidth]{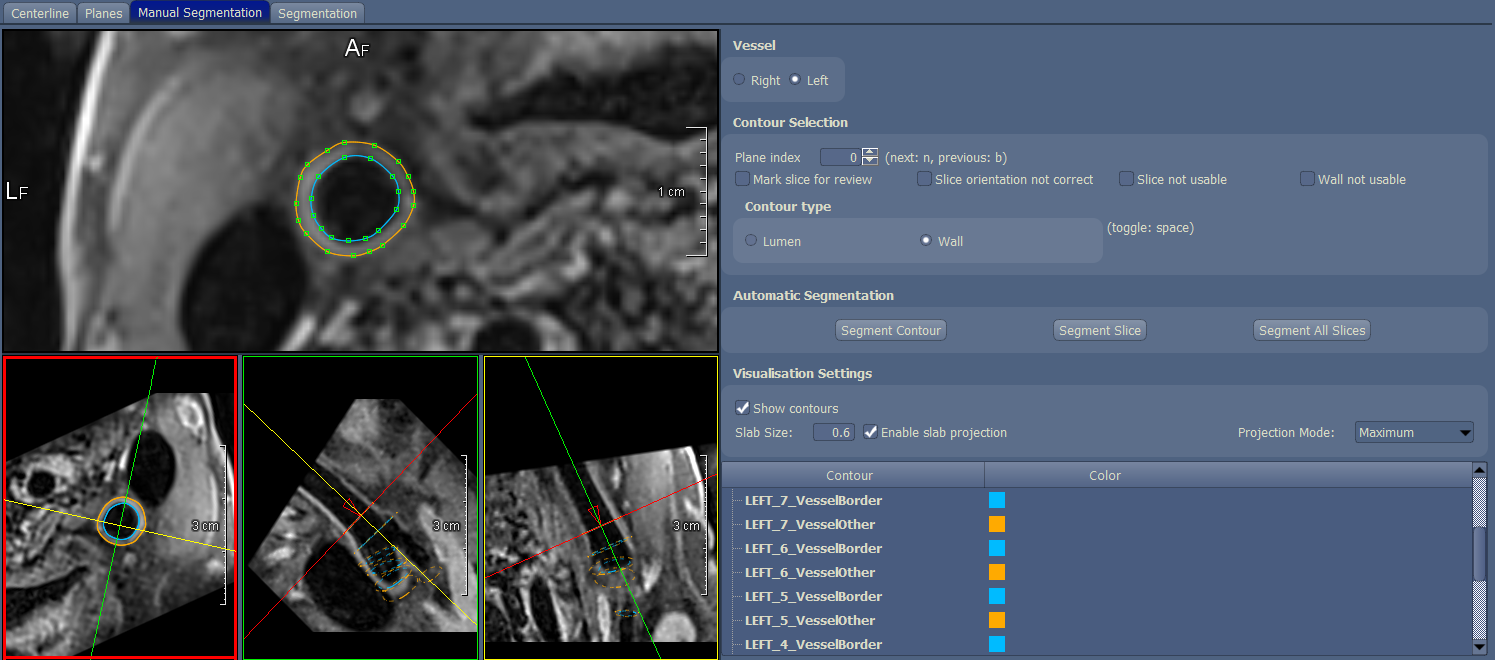}
  \caption{\label{fig:manual-segmentation-tab}%
           Tab for manual segmentation of the carotid artery in cross-sections that are perpendicular to the centerline. Automatic segmentations can be created and are transformed into contours. This allows for manual correction and good visibility of the BB-MRI data.}
\end{figure*}

For the segmentation of the carotid artery in perpendicular 2D cross-sections, the manual segmentation tab is used (Fiugre~\ref{fig:manual-segmentation-tab}). In the top left, the cross-section is shown and can be manually segmented by creating a lumen and a wall contour. To create a contour, seed points are manually placed, and the contour is given by a spline that connects them. The contour can be corrected by dragging or adding seed points.

The vessel wall can also be automatically segmented using a convolutional neural network and a post-processing step that transforms the network result into a contour with seed points. This allows a consistent visualization and correction of automatically created contours\cite{RHS*24}. 

Cross-sections with low signal-to-noise ratio or low contrast can be labeled as "not usable", and overlaying of contours can be disabled to allow an unbiased view of the MRI data. Three orthogonal views are shown in the lower left of the tab. They allow utilization of 3D context for the segmentation, and the segmentation along other planes can be seen.

\subsection{Visualization of 3D segmentation and VWT}
An automatic 3D segmentation of the carotid artery in BB-MRI is currently developed. To assess the correctness of the 3D segmentation, CaroTo uses a curved multiplanar reconstruction (MPR). The centerlines of the common and internal carotid arteries are used for the curved MPR, and the 3D segmentation result is overlayed as shown in Figure~\ref{fig:visualization}a. A small plaque in the posterior  can be seen on the left.

To visualize the distribution of wall thickness in 3D, CaroTo creates a 3D rendering showing the outer vessel wall as a transparent surface and the inner vessel wall color-coded with the VWT. Figure~\ref{fig:visualization}b shows this visualization for a carotid artery with a mild stenosis, and the increased wall thickness in the bifurcation area and in the posterior ICA can be easily seen.

\subsection{Visualization of 4D-Flow MRI}
We use animated pathlines to visualize the flow in the carotid artery as proposed by van Pelt et al. \cite{PBB*10}. We use color-coding to visualize the blood flow velocity, and different cross-sections of the carotid artery can be chosen as emitters for the pathlines. To focus the users attention on the relevant areas, the pathlines are masked with the 3D PC-MRA segmentation.

\begin{figure}[htb]
  \centering
  \includegraphics[width=1\linewidth]{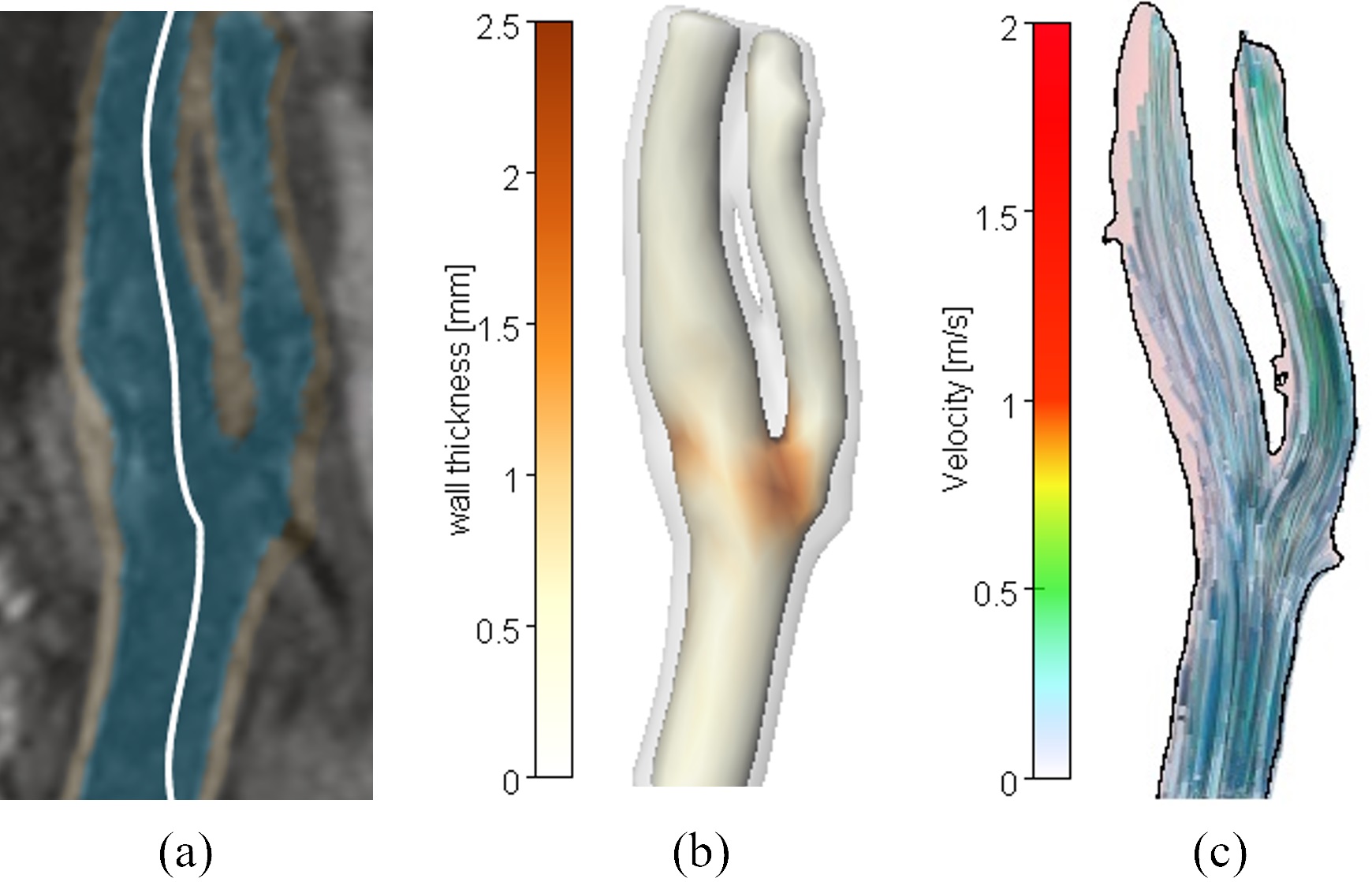}
  \caption{\label{fig:visualization}
           Visualization of (a) 3D segmentation on a curved MPR (b) 3D segmentation mask, color-coded with the VWT (c) Pathlines computed from 4D-Flow MRI, color-coded with the velocity}
\end{figure}

Figure~\ref{fig:visualization}b shows the visualization for a carotid artery with a mild stenosis. The pathlines do not show any vortices and no areas where the blood flows with high velocities above 1~$\frac{m}{s}$. The flow velocities in the external carotid artery (right) are higher than in the internal carotid artery (left).

\subsection{Conclusions}

We developed a tool for the assessment of carotid atherosclerosis in MRI data and are continuously adding new features like automatic segmentation and additional biomarkers. 

CaroTo is not publicly available due to licensing reasons. It can be requested from the authors at hinrich.rahlfs@dhzc-charite.de.

\section*{Acknowledgements} 
This work is funded by the German Research Foundation (GRK2260, BIOQIC)

\bibliographystyle{eg-alpha-doi} 
\bibliography{egbibsample}       

@article{CDM*19,
  title={Ischaemic stroke},
  author={Campbell, Bruce CV and De Silva, Deidre A and Macleod, Malcolm R and Coutts, Shelagh B and Schwamm, Lee H and Davis, Stephen M and Donnan, Geoffrey A},
  journal={Nature reviews Disease primers},
  volume={5},
  number={1},
  pages={70},
  year={2019},
  publisher={Nature Publishing Group UK London}
}

@article{SKK*20,
  title={Carotid geometry is an independent predictor of wall thickness--a 3D cardiovascular magnetic resonance study in patients with high cardiovascular risk},
  author={Strecker, Christoph and Krafft, Axel Joachim and Kaufhold, Lilli and H{\"u}llebrandt, Markus and Weber, Susanne and Ludwig, Ute and Wolkewitz, Martin and Hennemuth, Anja and Hennig, J{\"u}rgen and Harloff, Andreas},
  journal={Journal of Cardiovascular Magnetic Resonance},
  volume={22},
  pages={1--12},
  year={2020},
  publisher={Springer}
}

@article{RHS*24,
  title={Learning carotid vessel wall segmentation in black-blood MRI using sparsely sampled cross-sections from 3D data},
  author={Rahlfs, Hinrich and Hüllebrand, Markus and Schmitter, Sebastian and Strecker, Christoph and Harloff, Andreas and Hennemuth, Anja},
  journal={J Med Imaging (Bellingham)},
  volume={11},
  number={4},
  year={2024}
}

@article{PBB*10,
  title={Exploration of 4D MRI blood flow using stylistic visualization},
  author={Van Pelt, Roy and Besc{\'o}s, Javier Oliv{\'a}n and Breeuwer, Marcel and Clough, Rachel E and Gr{\"o}ller, M Eduard and ter Haar Romenij, Bart and Vilanova, Anna},
  journal={IEEE transactions on visualization and computer graphics},
  volume={16},
  number={6},
  pages={1339--1347},
  year={2010},
  publisher={IEEE}
}


\newpage
\end{document}